\documentclass[5p,twocolumn]{elsarticle}
\usepackage{graphicx,latexsym}
\usepackage{dcolumn}
\usepackage{amssymb,amsmath,bm}
\usepackage{subfigure}
\usepackage{braket}
\usepackage{siunitx}
\usepackage{array}
\usepackage{float}
\usepackage[nopar]{lipsum}
\usepackage{caption}
\usepackage{multirow}
\usepackage{bigstrut}
\usepackage[super]{nth}

\newcommand{\angstrom}{\text{\normalfont\AA}}
\usepackage{hyperref}
\hypersetup{
    pdfnewwindow=true,       % links in new window
    colorlinks=true,         % false: boxed links; true: colored links
    linkcolor=blue,          % color of internal links
    citecolor=blue,          % color of links to bibliography
    filecolor=magenta,       % color of file links
    urlcolor=black           % color of external links
}

\usepackage[normalem]{ulem}

\def\fig#1{Fig.\ \ref{#1}}
\def\tab#1{Tab.\ \ref{#1}}

\journal{}

\begin{document}

\begin{frontmatter}

%-----------------------------------------------------------------

\title{Role of interlayer spacing on electronic, thermal and optical properties of BN-codoped bilayer graphene:\break Influence of the interlayer and the induced dipole-dipole interactions}

\author[a1,a2]{Nzar Rauf Abdullah}
\ead{nzar.r.abdullah@gmail.com}
\address[a1]{Division of Computational Nanoscience, Physics Department, College of Science, 
             University of Sulaimani, Sulaimani 46001, Kurdistan Region, Iraq}
\address[a2]{Computer Engineering Department, College of Engineering, Komar University of Science and Technology, Sulaimani 46001, Kurdistan Region, Iraq}

\author[a1]{Hunar Omar Rashid}

\author[a4]{Chi-Shung Tang}
\address[a4]{Department of Mechanical Engineering,
	National United University, 1, Lienda, Miaoli 36003, Taiwan}

\author[a5]{Andrei Manolescu}
\address[a5]{Reykjavik University, School of Science and Engineering,
	Menntavegur 1, IS-101 Reykjavik, Iceland}

\author[a6]{Vidar Gudmundsson}
\address[a6]{Science Institute, University of Iceland, Dunhaga 3, IS-107 Reykjavik, Iceland}

%----------------------------------------------------------------

\begin{abstract}

We demonstrate that the electronic, thermal, and optical properties of a graphene bilayer with boron and nitrogen dopant atoms can be controlled by the interlayer distance between the layers in which the interaction energy and the van der Waals interaction between the dopant atoms play an essential role. We find a conversion of an AA- to an AB-stacked bilayer graphene caused by the repulsive interaction between dopant atoms. 
At a short interlayer distance, a strong repulsive interaction inducing a strong electric 
dipole moment of the dopant atoms is found. This gives rise to a breaking of the high symmetry, opening up a bandgap. 
Consequently, a considerable change in thermoelectric properties such as the Seebeck coefficient and the figure of merit are seen. The repulsive interaction is reduced by increasing the interlayer distance, and at a large interlayer distance the conversion process of the stacking order vanishes. 
A small bandgap is found leading to a low Seebeck coefficient and a figure of merit. 
For both short and large interlayer distances, a prominent peak in the optical response is found in the visible range and the peak position is inversely proportional to the interlayer distance.

\end{abstract}

\begin{keyword}
Thermoelectric \sep Bilayer graphene \sep DFT \sep Electronic structure \sep  and Optical properties
\end{keyword}

\end{frontmatter}

\section{Introduction} 

Physical properties of bilayer graphene (BLG) have drawn a great deal of experimental and theoretical attention due to interest in it's high potential electronics, thermal, and optical applications \cite{MCCANN2007110,  Gabor648, PhysRevB.95.085435}.
For instance, the electronic bandgap of BLG can be tuned via foreign dopant atoms or electric field that break the sub-lattice symmetry \cite{ABDULLAH2020100740, PhysRevLett.102.256405}. Consequently, the bandgap induces good thermal and optical properties of the system \cite{Ahn2018, Do2019}.
This has led to interest in BLG in photonics and optoelectronics, presented by its strong application possibilities ranging from solar cells and light-emitting devices to touch screens, photo-detectors and ultrafast lasers \cite{Bonaccorso2010, C4TA01047G}.

Several properties and applications of BLG are related to the interaction between the 
graphene layers. 
Obviously, the interlayer interaction energy is very useful for controlling electrical, thermal,
and mechanical properties of BLG \cite{doi:10.1021/nn1031017, Lui2011}. 
It has been shown that the interlayer equilibrium distance providing the interaction between two layers of graphene is $3.1\text{–}3.5$~$\angstrom$, with a binding energy of $60\text{–}72$~meV \cite{Chen2013}.
The interlayer interaction is defined by the interaction energy, the van der Waals (vdW), and the London dispersion forces that play a crucial role in many physical, chemical, and biological systems \cite{PhysRevB.74.205434, ARULSAMY2013}.
The best approach to describe the interlayer interaction is van der Waals Density Functional (vdW-DF) description. Generally, comparing with the traditional semilocal approach, the
generalized gradient approximation, GGA \cite{PhysRevB.33.8800, ABDULLAH2021114644}, and the local density approximation, LDA \cite{PhysRev.140.A1133, doi:10.1063/1.2822113}, it is found that the semiempirical vdW correction gives rise to minor changes of physical proprieties of a system \cite{PhysRevB.73.205101}.
The two approaches, LDA and GGA, are less accurate in describing layered surfaces of BLG or graphite. In addition, the interaction of atoms or surfaces at large separations is often not correctly described within LDA or GGA. This is attributed to the dispersion interaction between the electrons.

The interlayer vdW coupling of transition-metal atoms doped in BLG has been investigated and 
it has been shown that a mirror twin boundary modifies the interlayer vdW coupling in BLG leading to the development of local strain effective for few nanometers \cite{doi:10.1021/acsnano.7b09029}. The local strain thus influences local conductance and thermoelectric properties. Furthermore, some important variations in the complex dielectric function and related properties in the visible-light spectral region are found due to the presence of the interlayer vdW coupling \cite{ULIAN2021112978}.
On the other hand, it observed that the adhesion energy of graphene on SiC and h-BN is invariant under the dispersion corrections, whether described by the local, the semi-local, or the 
van der Waals interaction-corrected density functional theory (DFT) \cite{UKPONG20151}.

In the present work, we consider BN-codoped BLG with vdW interlayer interaction between the layers. 
The distance between layers is tuned and we find that the  electronic, thermal and optical properties of the system in the presence of vdW interlayer interaction are strongly influenced when the interlayer distance is less than $4.0 \, \angstrom$, but they are invariant when the interlayer distance is greater than $4.0 \, \angstrom$.

\section{Model and Method}

Our model consists of a $2\times2$ supercell BLG with B and N dopant atoms, where the B atom is doped in the top layer and the N atom is doped in the bottom layer.  To visualize the structures, we use the XCrySDen software \cite{KOKALJ1999176}. 
The calculations have been performed using the Quantum espresso (QE) package which uses plane waves as a basis set. The approach is based on an iterative solution of the Kohn-Sham equations of the DFT \cite{PhysRev.140.A1133}. 
In order to take into account the vdW interaction in our model, we consider the DFT-D technique and vdW-DF exchange–correlation functional \cite{Berland_2015}. We can thus take into account the long-range electron correlation. 
The Monkhorst–Pack scheme is used for sampling the Brillouin zone and plane-wave cutoff energy is set to $ \approx 1360$ eV.
For the geometric optimizations, a Gamma-centered $k$-mesh grid with $20\times20\times2$ is used and the structures are fully relaxed until the Hellmann-Feynman forces are smaller than $10^{-6}$~eV/$\angstrom$.
The thermal properties of the systems are calculated using the BoltzTrap software  \cite{madsen2006boltztrap-2}.
The optical spectra for the pure and the BN-codoped BLG structures
are calculated using a $80 \times 80 \times 2$ $k$-mesh grid of the  $\Gamma$ centering and setting Lorentzian broadening of $0.1$ eV.

 \section{Results and discussion}
 
\begin{figure}[htb]
 	\centering
 	\includegraphics[width=0.5\textwidth]{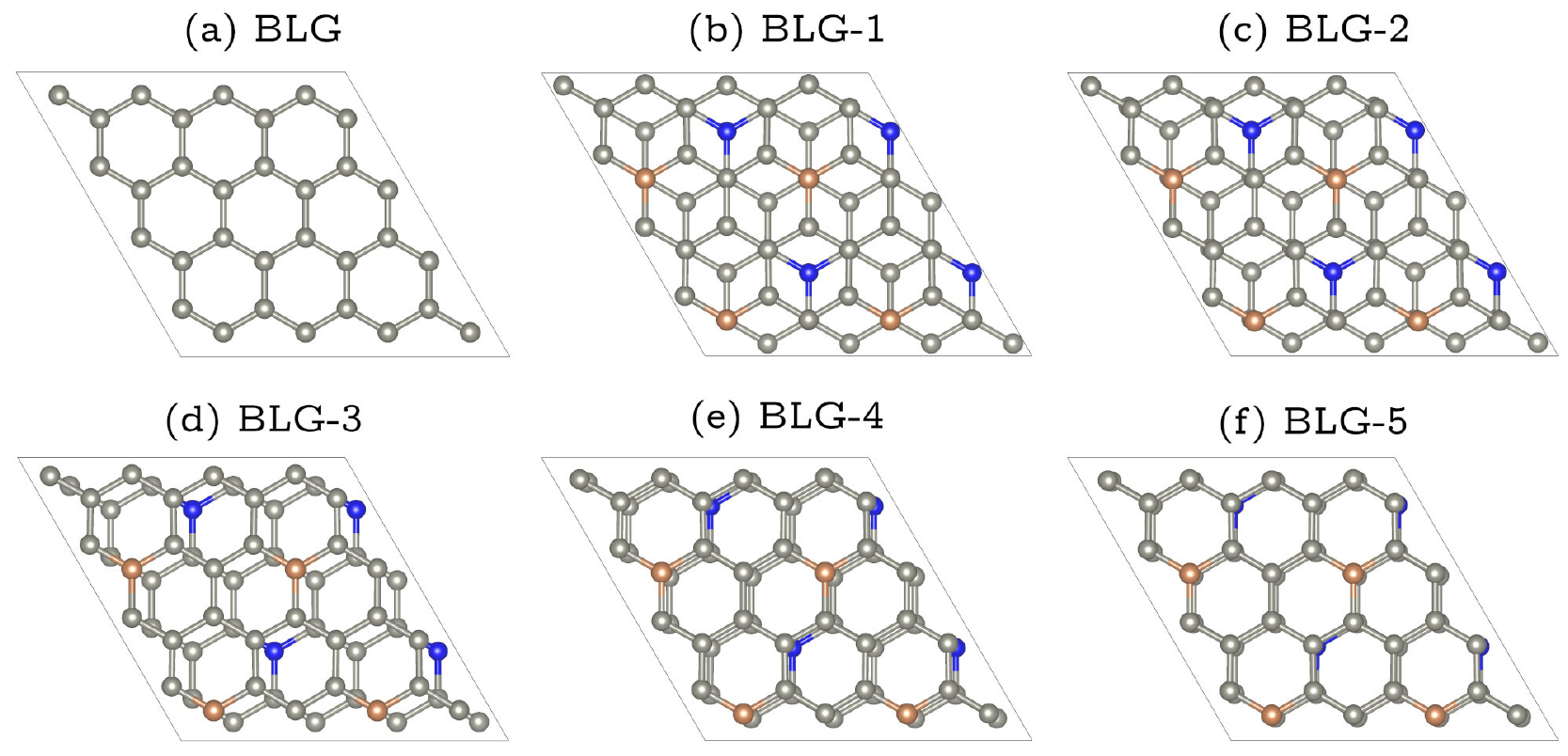}
 	\caption{Pure BLG (a) and BN-codoped BLG with the interlayer distance $d = 2.96$ (BLG-1), $3.5$ (BLG-2), $4.0$ (BLG-3), $4.5$ (BLG-4), 
 	$5.0 \, \angstrom $ (BLG-5). The C, B and N atoms are gray, orange, and blue colored, respectively.}
 	\label{fig01}
 \end{figure}

 In this section, we present the results for the electronic, optical and thermal properties of a pure and BN-codoped BLG with different interlayer distances.  Figure \ref{fig01} shows the pure BLG (a) and BN-codoped BLG (b-f) for different interlayer distances.
 The B atom is doped at a para-position of top layer while the N atom is doped at a ortho-position in the bottom layer \cite{ABDULLAH2020126350, ABDULLAH2020103282, https://doi.org/10.1002/andp.201500298}.
 The structural properties of our models are shown in \tab{table_one}. 
After fully relaxing the structures the distance between the layers, $d$, for pure AA-stacked BLG  is found to be $3.6 \, \angstrom$ which is in agreement with experimental observations \cite{PhysRevLett.99.216802, Chung2002}, and 
the interlayer distance of BLG with BN dopant atoms, identified as BLG-1, is 
$2.95 \, \angstrom$. 
 \begin{table}[h]
	\centering
	\begin{center}
		\caption{\label{table_one} Interlayer distance, $d$, distance between B and N atoms, $d_{\rm BN}$, interaction energy, $\Delta E$, and bandgap, $\Delta_g$, for pure BLG and BN-codoped BLG structures.}
		\begin{tabular}{|l|l|l|l|l|l|l|l|}\hline
			Structure	 & d ($\angstrom $)   & d$_{\rm BN}$ ($\angstrom $)   & $\Delta E$ (eV)    & $ \Delta_g$  (eV) \\ \hline
			BLG	    &  3.6    &   -----    &     -----    & 	0.0 	\\
			BLG-1   &  2.95   &   4.107    &     2.39     &   0.512	    \\
			BLG-2	&  3.5    &   4.6      &     2.12     &   0.24      \\
			BLG-3	& 4.0     &   4.9      &     1.13     &   0.13      \\   
			BLG-4	& 4.5     &   5.4      &     0.64     &  ---        \\
			BLG-5	& 5.0     &   5.8      &     0.23     &  ---        \\   \hline
	\end{tabular}	\end{center}
\end{table}
It is interesting to mention that the AA-stacking structure is converted to a AB-stacking one in BLG-1 (see \fig{fig01}(b)), which is due to the repulsive interaction between the B and N atoms as we recently reported \cite{ABDULLAH2020100740, nzar_2021}. The repulsive interaction between the B and N atoms in BLG-1 is $2.39$~eV and the distance between the B and N atom is found to be $4.107 \, \angstrom$ \cite{doi:10.1063/1.4742063, ABDULLAH2020126807}. 
The positive sign of the interaction energy, $\Delta E$, displays a repulsive interaction between the two atoms.

We {\em artificially} increase the distance between the layers of the BN-codoped BLG to $3.5$ (BLG-2), $4.0$ (BLG-3), $4.5$ (BLG-4), and $5.0  \, \angstrom$ (BLG-5), and then let the structures relax again, but keep the interlayer distance fixed for all structures mentioned above. So, we have the fully relaxed structures but with the selected interlayer distance. The maximum distance between the B and N atoms, d$_{\rm BN} = 5.8\, \angstrom$, is found for BLG-5.
The repulsive interaction between the B and N atoms is drastically decreased and the shift
in the stacking orientation gradually vanishes from the BLG-2 to the BLG-5 structure (see \fig{fig01}(c-f)). We find that the most stable structure among the BN-codoped BLG is the BLG-1 which has a minimum configuration energy. This is expected because a perfect AB-stacked BLG is more stable than BLG structures with other stacking types\cite{OuldNE2017, nzar_2021}.
It should be mentioned that the interaction between the B and N atoms in the presence 
of the vdW-DF is active up to $d_{\rm BN} \approx 5.0 \, \angstrom$ which is a new finding of our study. 
It was reported that the interaction strength is almost zero when the separation distance between the B and N atoms is greater than or equal to $4.0$~eV in the absence of the vdw-DF when the GGA or the LDA are considered \cite{doi:10.1063/1.4742063, ABDULLAH2020114556}.

\begin{figure}[htb]
	\centering
	\includegraphics[width=0.35\textwidth]{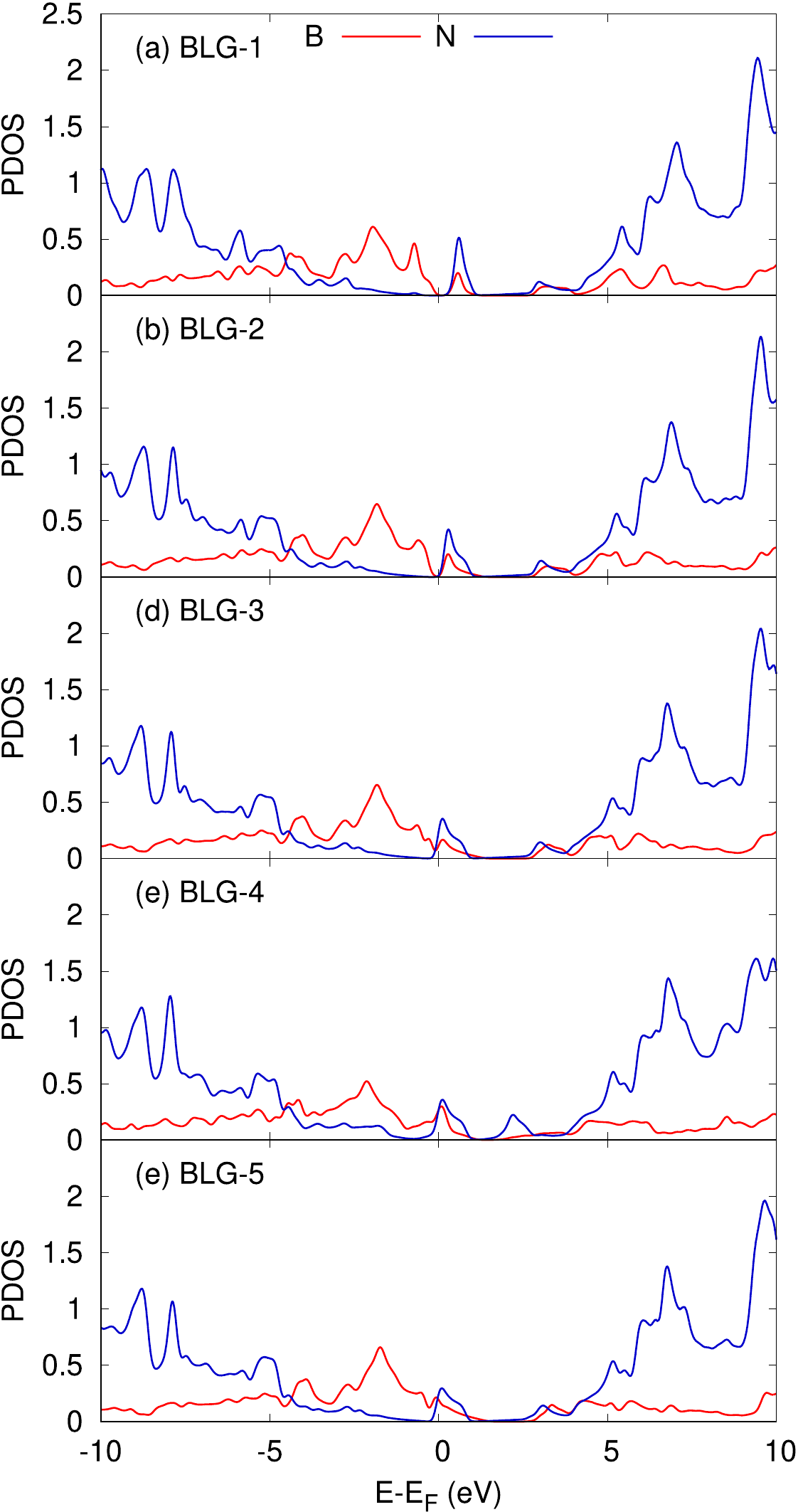}
	\caption{Partial density of state (PDOS) of the B (red) and the N (blue) atoms for the BLG-1 (a), BLG-2 (b), BLG-3 (c), BLG-4 (d), and BLG-5 (e) structures. The Fermi energy is set at $0$ eV.}
	\label{fig02}
\end{figure}

\begin{figure}[htb]
	\centering
	\includegraphics[width=0.5\textwidth]{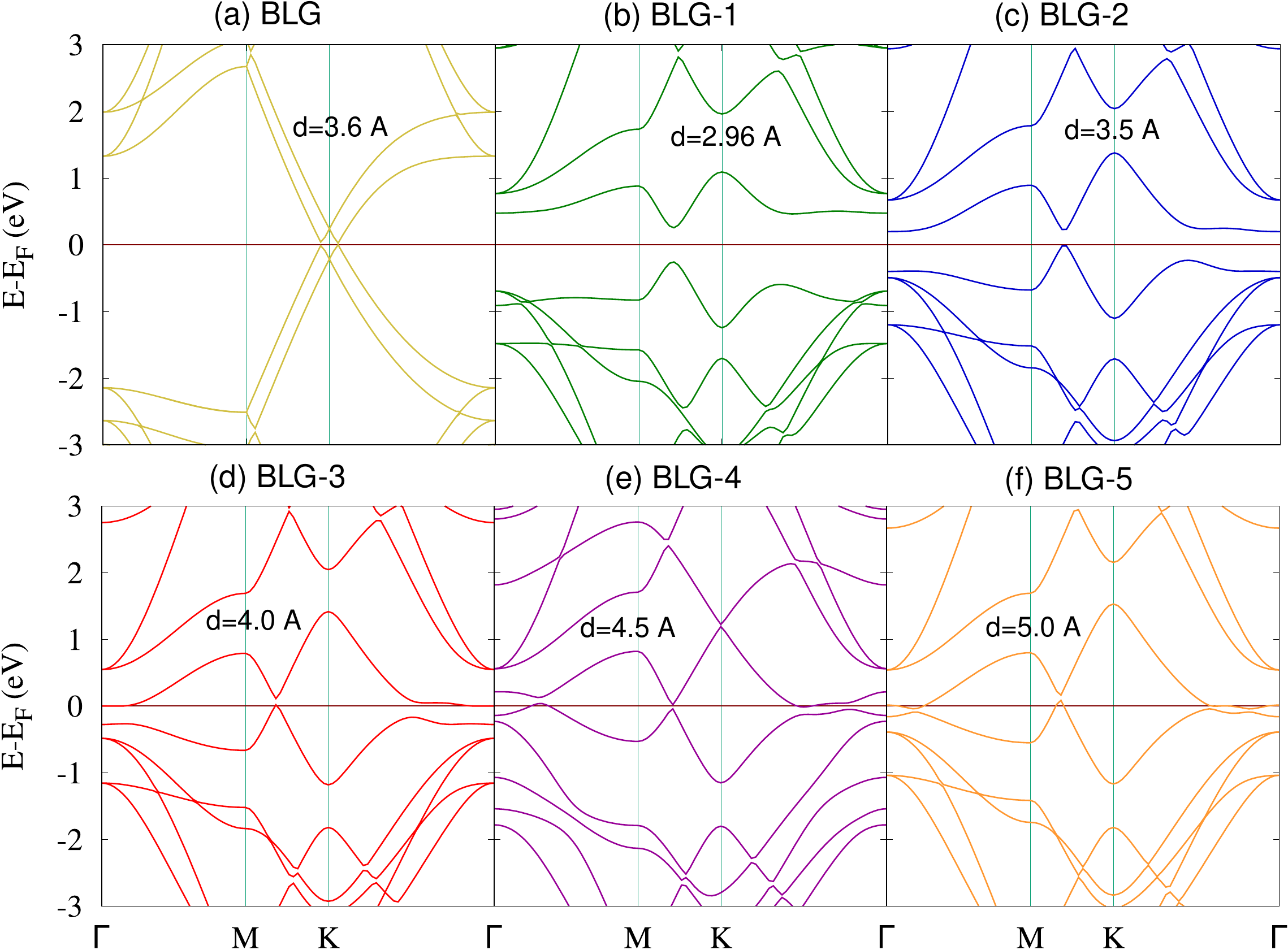}
	\caption{Electronic band structure of pure BLG (a), BLG-1 (b), BLG-2 (c), BLG-3 (d), BLG-4 (e), BLG-5 (f). The Fermi energy, $E_F$, is set at $0$ eV.}
	\label{fig03}
\end{figure}

In fact, the presence of the B and N atoms in the BLG structure leads to a charge transfer occurring from the N-doped layer to the B-doped layer as the electronegativity of an N atom is higher than that of a B atom. This redistribution of electronic charge in the BN-codoped BLG arises a dipole moment which generates an electric field between the layers along the direction perpendicular to both layers \cite{PhysRevB.93.205420}.
It has been reported that the electric dipole-dipole interaction becomes significant if the distance between the dipoles is less than $5 \, \angstrom$ \cite{wohlert2004range}. 
So, in addition to the repulsive interaction between the B and N atoms mentioned above, there is also an electric dipole-dipole interactions generated by the B and N atoms, which can be 
confirmed by the partial density of state (PDOS) shown in \fig{fig02}. 
One can see that the lowest unoccupied band belongs to the N-doped layer and the highest occupied band belongs to the B-doped layer around the Fermi energy. This can be realized from
the PDOS of a B atom which is mainly below the Fermi energy and the PDOS of a N atom mainly contributing above the $E_F$. 
The PDOS around the Fermi energy for both the B and N atoms are p$_z$-components indicating 
the $\pi$ state of a B atom and $\pi^*$ of the N atom. 
In addition, one can estimate the bandgap of each structure from the PDOS in which the overlap between the PDOS of the N and B is increased with the interlayer distance reflecting the bandgap shown in \fig{fig03}.

In \fig{fig03}, the dispersion energy is plotted for the undoped BLG (a), and the BN-doped BLG with the interlayer distances $2.95$ (b), $3.5$ (c), $4.0$ (d), $4.5$ (e), and $5.0 \, \angstrom$ (f).
The energy dispersion of a pure BLG (a) is re-plotted here in order to compare it with the band structures of the BN-codoped BLG. The Dirac point is found at the symmetric K-point for the undoped BLG in the hexagonal Brillouin zone, and it does not exist anymore at the K-point in the presence of the BN dopant atoms (see \fig{fig03}(b-f)). 
It can be seen that the bandgap, $\Delta_g$, is located away from K-point, along the $\Gamma$-K direction, and the generated bandgap is direct for the BLG-1 and BLG-2, but indirect for the BLG-3.  The bandgap values of all the structures are presented in \tab{table_one}. 
We confirm that the interlayer distance or the distance between the B and N atom impurities has a major influence on the band structure. 
The bandgap of BLG-1 is maximum and it has almost vanished for the BLG-5 along the M-K path as the interlayer distance is increased.  We should mentioned that the energy dispersion of the BN-codoped BLG structures are qualitatively in good agreement with earlier results \cite{PhysRevB.93.205420}.   
\lipsum[0]
\begin{figure*}[]
	\centering
	\includegraphics[width=0.7\textwidth]{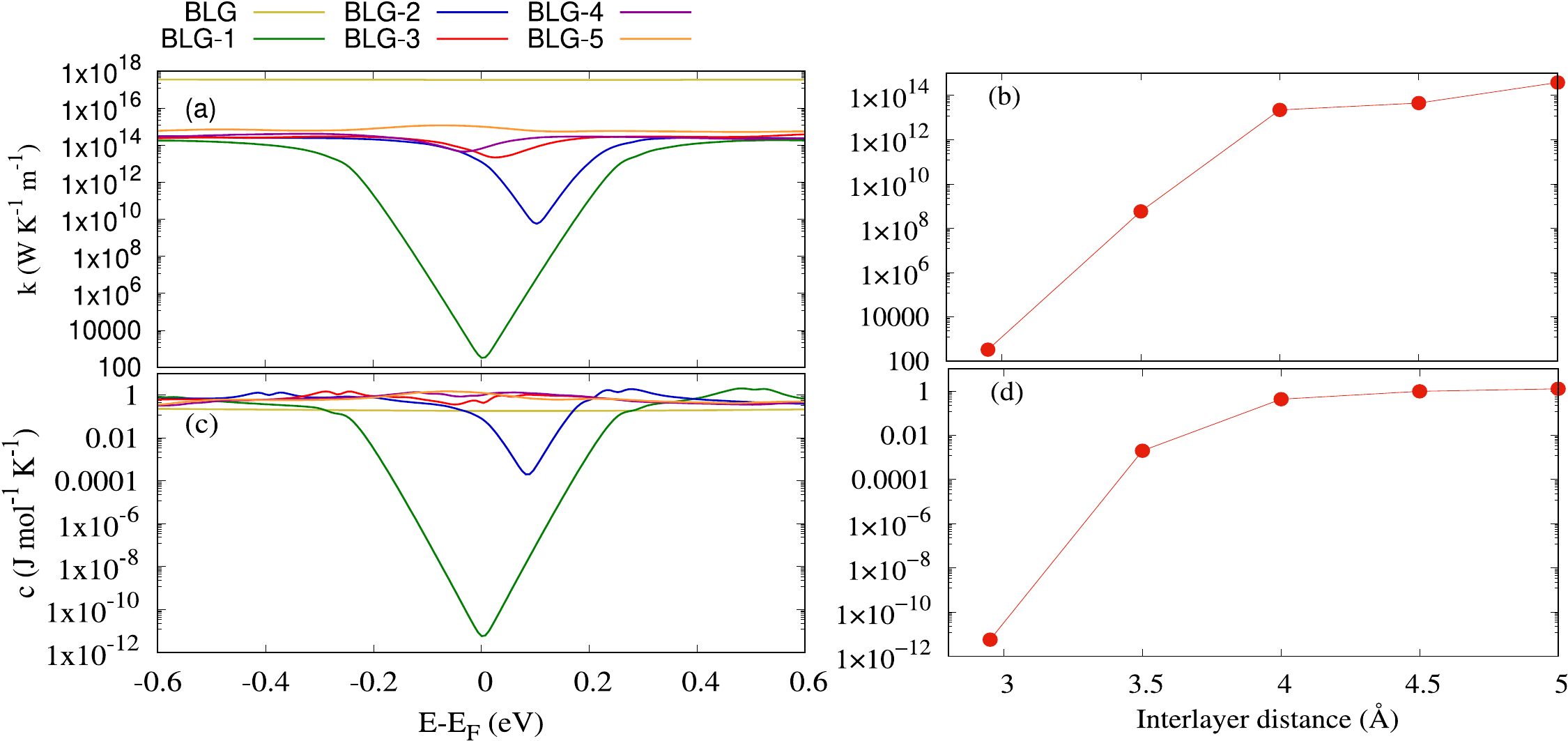}
	\caption{Electronic thermal conductivity (a,b), and specific heat (c,d) of the pristine and BN-codoped BLG structures at $100$~K.}
	\label{fig04}
\end{figure*} 
In addition, the energy dispersion around K-point demonstrates two ``mexican hats'' which is similar to the band structure of BLG subject to a perpendicular electric field \cite{PhysRevLett.99.216802, Chuang2013}, where the external electric field breaks the sub-lattice symmetry of the BLG.
The symmetry breaking of our systems is partly caused by the intrinsic electric field generated by the B-N dipole. The generated electric field in our systems is decreased as the interlayer distance increases. The smaller the electric field, the less the sub-lattice symmetry breaking emerges.
This leads to an almost pure AA stacking of BLG-5 as is the case for a pure BLG. 
Consequently, a smaller bandgap is obtained for the BLG-5 along the M-K path.

The electronic thermal properties are dominant in the system in the temperature range 
of $20\text{-}160$~K, where the lattice thermal motion has a minor contribution to 
the conductance \cite{PhysRevB.87.241411}.
To study the electronic thermal properties, we use the BoltzTraP software which is a package to calculate the semi-classic transport coefficients. The code uses a mesh of band energies and is interfaced to the QE package.

In order to show the thermoelectric performance of materials and devices, we should have a high  figure of merit or a power factor \cite{ABDULLAH2020126578, Abdullah2019}. It is well known that a high figure of merit, $ZT = (S^2 \sigma/k) T$, requires a low electronic thermal conductivity, $k$, and a high Seebeck coefficient, $S$, and electrical conductivity, $\sigma$.  The electronic thermal conductivity (a) and the specific heat (c) are shown in \fig{fig04}, where the temperature is assumed to be $100$~K \cite{RASHID2019102625}.
One can see that the lack of thermal electrons in the bandgap range inevitably leads to a decrease in the electronic thermal conductivity. Therefore, when the bandgap is larger, the reduced electronic thermal conductivity is higher, and the thermal conductivity is lower around 
the Fermi energy \cite{XU2020103015}. The same scenario applies to the specific heat. So, the lowest electronic thermal conductivity and specific heat are found for BLG-1 which has the 
largest bandgap among the structures here.
In addition, the electronic thermal conductivity (b), and specific heat (d) as a function of interlayer distance are presented in \fig{fig04}. We remember that the bandgap is inversely propositional to the interlayer distance of the BN-codoped BLG. 
The larger the interlayer distance, the smaller the bandgap is leading to a high electronic 
thermal conductivity and specific heat (see \fig{fig04}(b,d)).  

\begin{figure}[htb]
	\centering
	\includegraphics[width=0.4\textwidth]{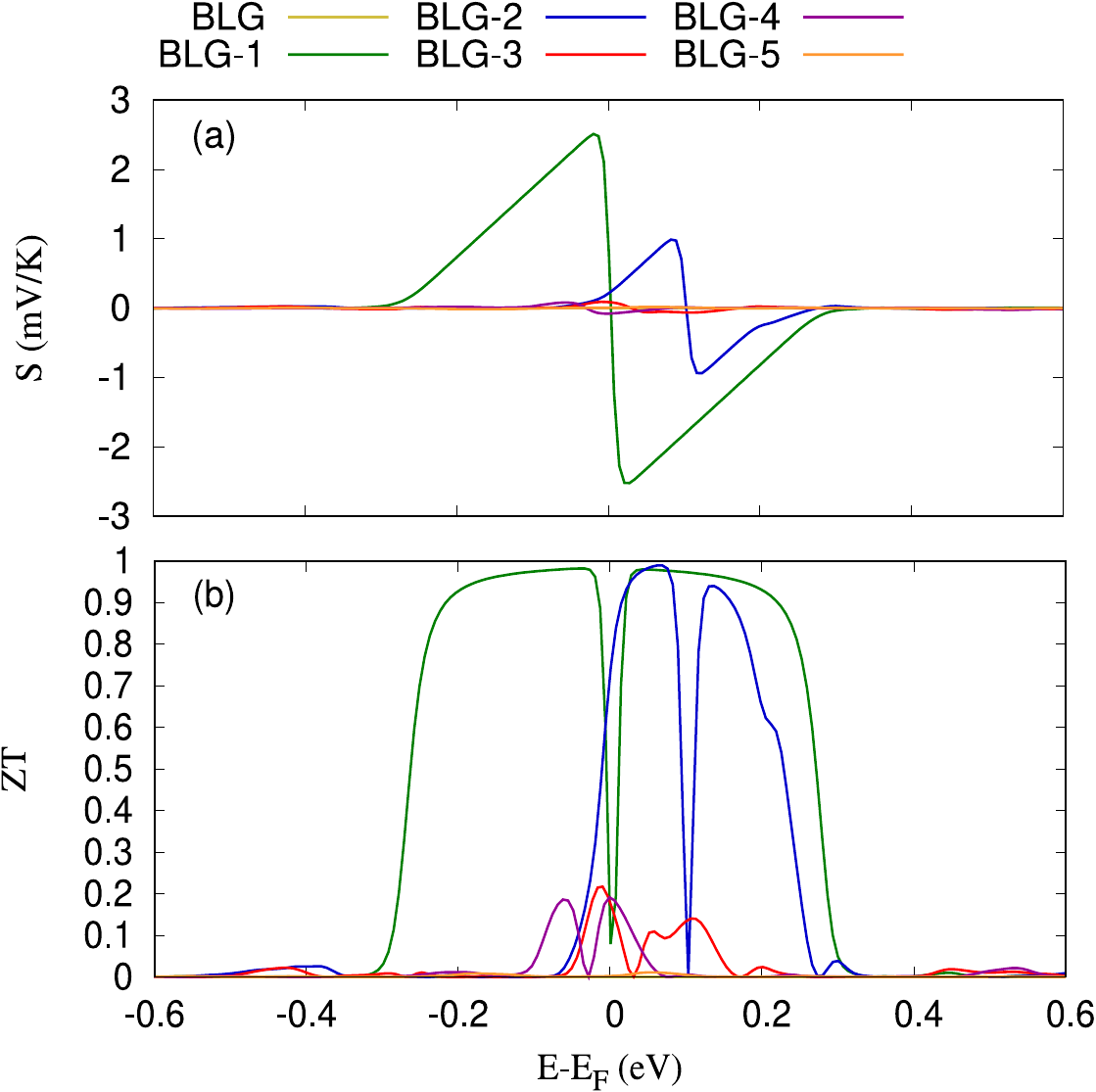}
	\caption{Seebeck coefficient, $S$, (a) and figure of merit, $ZT$, (b) of the pristine and BN-codoped BLG structures at $100$~K.}
	\label{fig05}
\end{figure}

The Seebeck coefficient, $S$, and the figure of merit, $ZT$, are displayed in \fig{fig05}. 
The symmetry between the valence and the conduction bands and the zero bandgap of pristine BLG, (see \fig{fig03}(a)), give rise a low Seebeck coefficient, and as a result a small $ZT$. The redistribution of charges due to the B and N dopant atoms breaks the sub-lattice symmetry and opens up the bandgap of BN-codoped BLG (see \fig{fig03}(b-e)). We therefore obtain the maximum $S$ for the the structure with the largest bandgap, BLG-1. The low electronic thermal conductivity and the high Seebeck coefficient of BLG-1 give rise the maximum $ZT$.   
The $S$ and $ZT$ gradually decrease with increasing interlayer distance until very small $S$ and $ZT$ are found for BLG-5, which has a very small bandgap along the M-K path. We therefore conclude that the BLG-1 is the best thermoelectric material among all considered structures here, and one can control the theroelectric properties by controlling the interlayer distance.

\begin{figure}[htb]
	\centering
	\includegraphics[width=0.4\textwidth]{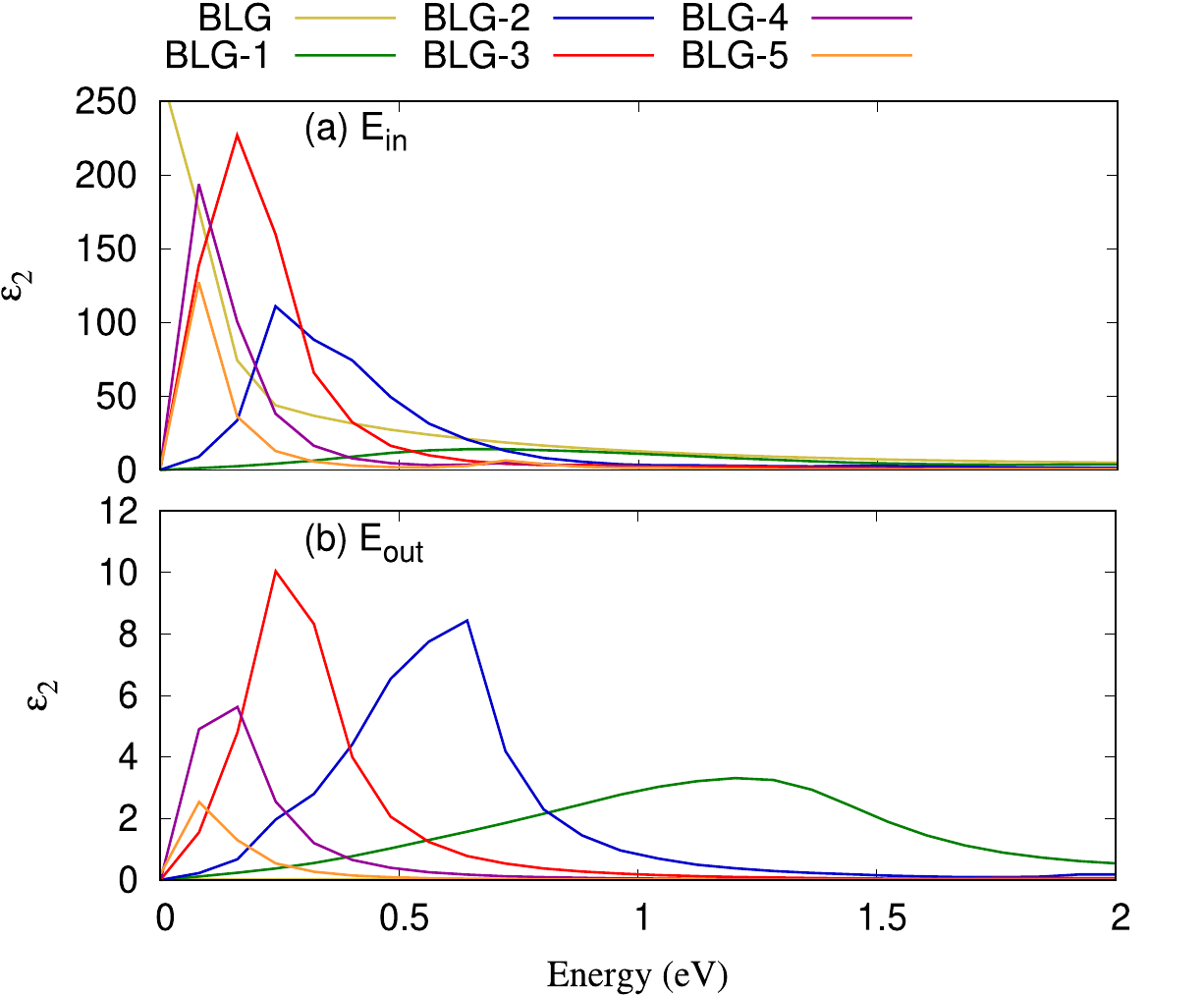}
	\caption{Imaginary part of the dielectric function in the case of parallel, E$_{\rm in}$, (a) and perpendicular, E$_{\rm out}$, (b) electric field.}
	\label{fig06}
\end{figure}

Since the interlayer interactions as well as the interlayer distances in BN-codoped BLG change the band structures from two pairs of linear bands to single bands, the optical transitions are expected to be modified.
To study the optical characteristics of systems, we calculated the dielectric function which indicates the linear response of systems subjected to the electromagnetic radiation.
Figure \ref{fig06} shows the imaginary part of dielectric function estimating the absorption spectra  of the structures when parallel (a), E$_{\rm in}$, or perpendicular (b), E$_{\rm out}$, electric field is applied.  
It has been shown that a red shift in $\varepsilon_2$ of BN-codoped BLG occurs for both the $\pi\text{-}\pi^*$ and the $\sigma\text{-}\sigma^*$ transitions in the case of $E_{\rm in}$, and $\sigma\text{-}\pi^*$ and $\pi\text{-}\sigma^*$ transitions for $E_{\rm out}$. In addition a high intensity peak in the optical response in the visible range is found \cite{ABDULLAH2020100740, ABDULLAH2020114221}. 
In order for BLG based optoelectronic devices to be useful, it is beneficial if they can be tailored to absorb in a specific wavelength region of the spectra.  
Here, we only show the peaks appear due to the presence of the B and N dopants in the low energy range of electromagnetic radiation from $0$ to $2.0$~eV.
It is interestingly found that the interlayer distance can tune the optical transition in the visible region (see \fig{fig06}). The peaks in $\varepsilon_2$ in the low energy ranges indicate the $\pi\text{-}\pi^*$ transitions in the case of E$_{\rm in}$, while the peaks for E$_{\rm out}$ manifest the $\sigma\text{-}\pi^*$ transitions. In the case of E$_{\rm in}$, 
the peak positions reflect the bandgap of the structures. The peak position is shifted towards the low energy as the interlayer distance is increased indicating the smaller bandgap at larger interlayer distance.
Furthermore, in the presence of E$_{\rm out}$ the peak position indicating the $\sigma-\pi^*$ transition is shifted towards lower energy because the energy spacing between the $\sigma$ and the $\pi^*$ bands is decreased with increasing interlayer distances.

\begin{figure}[htb]
	\centering
	\includegraphics[width=0.45\textwidth]{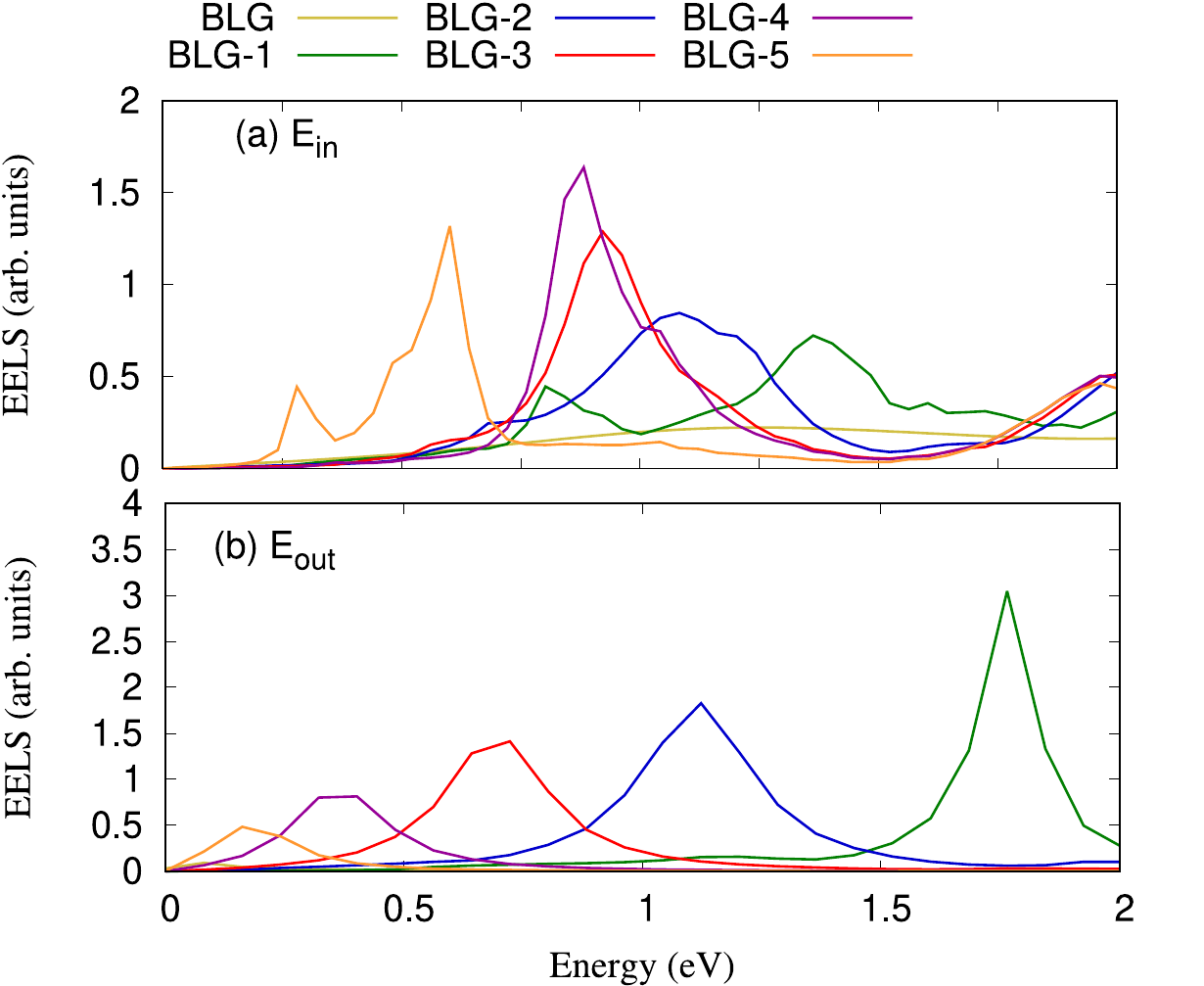}
	\caption{Electron energy loss spectrum (EELS) in the case of parallel, E$_{\rm in}$, (a) and perpendicular, E$_{\rm out}$, (b) electric field.}
	\label{fig07}
\end{figure}

The electron energy loss spectrum (EELS) that describes inelastic scattering of fast electrons in the structures is presented in \fig{fig07} for E$_{\rm in}$ (a), and E$_{\rm out}$ (b). The EELS is also useful for understanding the screened excitation spectra, especially the collective excitations. 
In the case of pure BLG, EELS exhibits no prominent peak due to the low density of state, 
but the BN dopants induces 
prominent peaks with the peak position depending on the interlayer distance.
The peak position of the EELS is inversely proportional to the interlayer distance for both E$_{\rm in}$ and E$_{\rm out}$. Our results for the EELS are qualitatively agree very well with Chuang et al.\ \cite{Chuang2013}, where the BLG is subjected to external electric field, and the peak position is shifted to a lower energy as the strength of external electric field is decreased.

\section{Conclusions}

In summary, we have calculated in detail the interaction between two graphene layers with B and N dopant atoms using density function theory and taking into account the van der Waals interaction. 
The distance between the layers depends on the positions of the dopants.
In these calculations, the Boltzmann equation has been used to study thermoelectric properties, and the independent particle approximation allowing us to describe single-particle excitations has been used for obtaining the optical response of the system.
At small interlayer distance, the charge transfers from the N to the B atoms is expected inducing a high dipole moment and generating an electric field. As a result, a strong interlayer interaction is observed leading to  improved thermal properties in the system due to the opening up of a bandgap.
Furthermore, a low interlayer interaction is found for larger interlayer distances reducing the thermal and optical response of the system.

\section{Acknowledgment}
This work was financially supported by the University of Sulaimani and 
the Research center of Komar University of Science and Technology. 
The computations were performed on resources provided by the Division of Computational 
Nanoscience at the University of Sulaimani. C.-S.T. acknowledges financial support from
Ministry of Science and Technology in Taiwan under grant No. MOST 109-2112-M-239-003.  
 
%\section{References}

%\bibliographystyle{elsarticle-num} 
%\bibliography{Ref_2.bib}

\end{document}